\documentclass[
  aip,
  pop,
  amsmath,amssymb,
  reprint,
]{revtex4-1}
\usepackage{graphicx,color}
\usepackage{amsmath}
\usepackage{natbib}
\usepackage{epsfig}
\begin{document}

\title{Gr\"uneisen parameter for strongly coupled Yukawa systems}

\author{Sergey A. Khrapak}
\affiliation{Aix Marseille University, CNRS, PIIM, Marseille, France; \\
Institut f\"ur Materialphysik im Weltraum, Deutsches Zentrum f\"ur Luft- und Raumfahrt (DLR), Oberpfaffenhofen, Germany; \\
Joint Institute for High Temperatures, Russian Academy of Sciences, Moscow, Russia}

\date{\today}

\begin{abstract}
The Gr\"uneisen parameter is evaluated for three-dimensional Yukawa systems in the strongly coupled regime. Simple analytical expression is derived from the thermodynamic consideration and its structure is analysed in detail. Possible applications are briefly discussed.     
\end{abstract}

\maketitle

\section{Introduction}

An equation of state (EoS) in the form of a relation between the pressure and internal energy of a substance (often referred to as the Gr\"uneisen or Mie-Gr\"uneisen equation) has been proven very useful in describing condensed matter under extreme conditions. Central to this form of EoS is the Gr\"uneisen parameter, whose thermodynamic definition is~\cite{Arp1984,Mausbach2016} 
\begin{equation}
\gamma_{\rm G}= V\frac{(\partial P/\partial T)_V}{(\partial E/\partial T)_V}=\frac{V}{C_V}\left(\frac{\partial P}{\partial T}\right)_V,
\end{equation}
where $V$ is the system volume, $P$ is the pressure, $T$ is the temperature, $E$ is the internal energy, and $C_V=(\partial E/\partial T)_V$ is the specific heat at constant volume. Under the assumption that $\gamma_{\rm G}$ is independent of $P$ and $E$ one can write~\cite{Arp1984,HummelPRB2015}
\begin{equation}
PV=\gamma_{\rm G}(\rho)E+C(\rho)V,
\end{equation} 
where $C(\rho)$ is the ``cold pressure'', which depends only on the density $\rho=N/V$. 

Gr\"unesein parameter depends considerably on the substance in question as well as on the thermodynamic conditions (location on the corresponding phase diagram). In most metals and dielectrics in the solid phase, $\gamma_{\rm G}$ is in the range from $\simeq 1$ to $\simeq 4$.~\cite{Arp1984} For fluids it is usually somewhat smaller, typically ranging from $\simeq 0.2$ to $\simeq 2$.~\cite{Arp1984} The focus of this paper is on Yukawa model systems, which are often applied as a first approximation to complex (dusty) plasmas, representing a collection of highly charged particles immersed in a neutralizing environment.~\cite{IvlevBook,FortovUFN2004,FortovPR2005,ShuklaRMP2009,CPBook,
MorfillRMP2009} In the context of complex plasmas, the Gr\"unesein parameter can be useful in describing shock wave phenomena observed in various complex plasma experiments.~\cite{SamsonovPRL2004,FortovPRE2005,HeinrichPRL2009,SaitouPRL2012,OxtobyPRL2013,
UsachevNJP2014} Therefore, it is desirable to have a practical approach allowing to estimate the Gr\"uneisen parameter and related quantities under different experimental conditions (an attempt to estimate $\gamma_{\rm G}$ has been previously reported in Ref.~\onlinecite{UsachevNJP2014}). In this paper we evaluate Gr\"uneisen parameter for strongly coupled three-dimensional (3D) one-component Yukawa systems.   

To be precise, Yukawa systems studied in this work represent a collection of point-like charged particles, which interact via the pairwise repulsive potential of the form
\begin{equation}\label{Yukawa}
V(r)= (Q^2/r)\exp(-r/\lambda),
\end{equation}
where $Q$ is the particle charge (assumed constant), $\lambda$ is the screening length, and $r$ is the distance between a pair of particles. Thermodynamics of considered Yukawa systems is fully characterized by the two dimensionless parameters. The first is the coupling parameter, $\Gamma=Q^2/aT$, where $a=(4\pi \rho/3)^{-1/3}$ is the characteristic interparticle separation (Wigner-Seitz radius) and $T$ is the temperature (in energy units). The second is the screening parameter, $\kappa=a/\lambda$. In the limit $\kappa\rightarrow 0$, the interaction potential tends to the unscreened Coulomb form, and Yukawa systems approach to the one-component-plasma (OCP).~\cite{Baus1980} Note, however, that in the OCP limit a uniform neutralizing background should be applied to keep the thermodynamic quantities finite. Thermodynamic properties of Yukawa systems received considerable attention. In particular, accurate data for the internal energy and compressibility obtained using Monte Carlo (MC) and molecular dynamics (MD) numerical simulations have been tabulated for a wide (but discrete) range of state variables $\Gamma$ and $\kappa$.~\cite{MeijerJCP1991,TejeroPRA1992,FaroukiJCP1994,HamaguchiPRE1997,
CaillolSP2000} Various integral theory approaches to the equation of state have also been used to describe strongly coupled Yukawa systems.~\cite{FaussurierPRE2004,ToliasPRE2014,ToliasPoP2015} Recently, a shortest-graph method has been applied to accurately describe thermodynamics of Yukawa crystals.~\cite{YurchenkoJCP2014,YurchenkoJCP2015}  

Simple and reliable analytical expressions for the energy and pressure of strongly coupled Yukawa fluids have been proposed in Refs.~\onlinecite{KhrapakPRE02_2015,KhrapakJCP2015}. These expressions are based on the Rosenfeld-Tarazona (RT) scaling~\cite{RT1,RT2} of the thermal component of the excess internal energy when approaching the freezing transition. These expressions demonstrate  relatively good accuracy~\cite{KhrapakPRE02_2015,KhrapakJCP2015} and are very convenient for practical applications. In this paper they are employed to estimate the Gr\"uneisen parameter of strongly coupled 3D Yukawa fluids. In this way very simple analytical expressions are obtained and analysed.    

\section{Thermodynamic properties}\label{Thermo}

The total system energy $E$ and pressure $P$ are the sums of kinetic and potential contributions. For 3D systems we can write
\begin{eqnarray}
E=\frac{3}{2}NT+U=\frac{3}{2}NT+NTu_{\rm ex}, \\
PV= NT+W=NT+NTp_{\rm ex},
\end{eqnarray}
where $U$ is the potential energy and $W$ is the configurational contribution to the pressure or virial. These are expressed in terms of conventional reduced (dimensionless) excess energy $u_{\rm ex}$ and excess pressure $p_{\rm ex}$, respectively. 

It should now be briefly reminded how the excess energy $u_{\rm ex}$  and pressure $p_{\rm ex}$ of one-component Yukawa fluids can be evaluated. We only provide the expressions required in subsequent calculations, further details can be found in Refs.~\onlinecite{KhrapakPRE02_2015,KhrapakJCP2015,KhrapakPPCF2016}.
The reduced excess energy of a strongly coupled Yukawa fluid can be approximated with a good accuracy by the expression
\begin{equation}\label{uEx}
u_{\rm ex}=M_{\rm f}\Gamma+\delta\left(\Gamma/\Gamma_{\rm m}\right)^{2/5}.
\end{equation}
Here the first term corresponds to the static energy contribution within the ion sphere model (ISM).~\cite{RT2,ISM} The quantity $M_{\rm f}$ is referred to as the {\it fluid} Madelung constant~\cite{RT2} and is given by 
\begin{equation}\label{M_f}
M_{\rm f}(\kappa) = \frac{\kappa(\kappa+1)}{(\kappa+1)+(\kappa-1)e^{2\kappa}}.
\end{equation}
The second term in Eq.~(\ref{uEx}) is the thermal contribution to the excess energy, which scales universally with respect to $\Gamma/\Gamma_{\rm m}$, where $\Gamma_{\rm m}$ is the coupling parameter at the fluid-solid (freezing) phase transition. This scaling holds for various soft repulsive particle systems, including the present case of Yukawa repulsion, provided the screening is not too strong.~\cite{RT2}  Regarding the dependence $\Gamma_{\rm m}(\kappa)$, it can be well described by a simple approximation~\cite{VaulinaJETP2000,VaulinaPRE2002}
\begin{equation}\label{melt}
\Gamma_{\rm m}(\kappa)\simeq \frac{172 \exp(\alpha\kappa)}{1+\alpha\kappa+\frac{1}{2}\alpha^2\kappa^2},
\end{equation}
where the constant $\alpha=(4\pi/3)^{1/3}\simeq 1.612$ is the ratio of the mean interparticle distance $\Delta=\rho^{-1/3}$ to the Wigner-Seitz radius $a$. The value of the constant $\delta$ in Eq.~(\ref{uEx}) is $\delta=3.1$, as suggested in Ref.~\onlinecite{KhrapakJCP2015}.
 
Using this approximation for the excess energy, the reduced pressure can be readily obtained as~\cite{KhrapakPRE02_2015}
\begin{equation}
p_{\rm ex} = p_0+\frac{\delta}{3}\left(\frac{\Gamma}{\Gamma_{\rm m}}\right)^{2/5}f_{\rm Z}(\alpha\kappa).
\end{equation}
Here $p_0$ is the static component of the pressure (associated with the static component of the internal energy)
\begin{equation}\label{p_0}
p_0= \frac{\kappa^4\Gamma}{6[\kappa{\rm cosh}(\kappa)-{\rm sinh}(\kappa)]^2},
\end{equation}
and the function $f_{\rm Z}$ is defined as
\begin{equation}\label{fZ}
f_{\rm Z}(x)=\frac{2+2x+x^2+x^3}{2+2x+x^2}.
\end{equation} 
The model described by Eqs.~(\ref{uEx}) - (\ref{fZ}) demonstrated excellent performance~\cite{KhrapakPRE02_2015,KhrapakJCP2015} in the regime $\kappa\lesssim 5$ and $\Gamma/\Gamma_{\rm m}\gtrsim 0.1$, which will be considered in this work.

\section{Relations between pressure and energy}

\subsection{Excess pressure-to-energy ratio}

\begin{figure}
\includegraphics[width=8 cm]{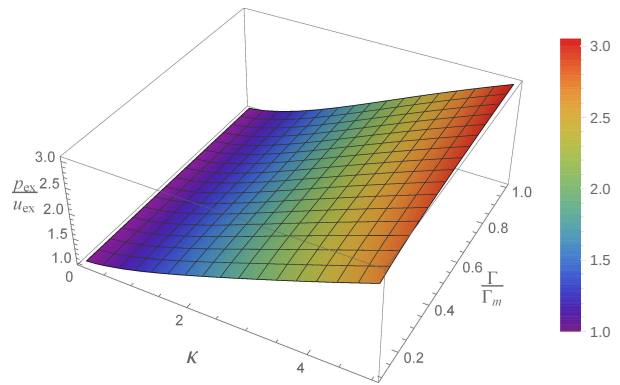}
\caption{(Color online) Plot of the ratio of the  excess pressure to the excess energy, $p_{\rm ex}/u_{\rm ex}$, on the plane of Yukawa systems state variables $\kappa$ and $\Gamma/\Gamma_{\rm m}$}
\label{Fig1}
\end{figure}

Using the approximation of Eqs.~(\ref{uEx}) - (\ref{fZ}), important  relationships between the pressure and internal energy of Yukawa fluids can be investigated. We start with evaluating simply the ratio of the virial $W$ to the potential energy $U$, which is equal to the ratio $p_{\rm ex}/u_{\rm ex}$. This ratio has been previously evaluated for 2D Yukawa fluids.~\cite{FengPoP2016,KryuchkovSM2016} The calculation for 3D Yukawa fluids, using the thermodynamic functions described above, is presented in Figure~\ref{Fig1}. We note that the excess pressure-to-excess energy ratio is not very sensitive to the reduced coupling parameter $\Gamma/\Gamma_{\rm m}$. On the other hand, the ratio exhibits strong dependence on the screening parameter $\kappa$ (it increases with $\kappa$).

\subsection{OCP limit}

An important observation in Fig.~\ref{Fig1} is that $p_{\rm ex}/u_{\rm ex}\rightarrow 1$ as $\kappa\rightarrow 0$. At first glance, this seems perhaps counter-intuitive, because one would naturally expect $p_{\rm ex}/u_{\rm ex}=1/3$ as in the OCP limit in 3D. We remind, that for inverse-power-law (IPL) interactions of the form $V(r)\propto r^{-n}$ in 3D, a general relationship $p_{\rm ex}=\tfrac{n}{3}u_{\rm ex}$ holds ($n$ is referred to as the IPL exponent). The difference should be attributed to the presence of the uniform neutralizing background in the OCP limit, which is absent in one-component Yukawa systems. Let us prove this mathematically. In the limit of very soft interaction, the energy and pressure at strong coupling ($\Gamma\gg 1$) are dominated by their static contributions. The series expansion of the fluid Madelung energy [Eq.~(\ref{M_f})] and the corresponding static pressure [Eq.~(\ref{p_0})] in the limit $\kappa\rightarrow 0$ yield
\begin{displaymath}
M_{\rm f}(\kappa)\Gamma\simeq -\frac{9\Gamma}{10}+\frac{\kappa\Gamma}{2}+\frac{3\Gamma}{2\kappa^2}+{\mathcal O}(\kappa^2\Gamma),
\end{displaymath} 
and
\begin{displaymath}
p_0(\kappa)\simeq -\frac{3\Gamma}{10}+\frac{3\Gamma}{2\kappa^2}+{\mathcal O}(\kappa^2\Gamma).
\end{displaymath}
In the absence of explicit thermodynamic contribution from the neutralizing medium (that is for one-component Yukawa systems), both $M_{\rm f}$ and $p_0$ are divergent at $\kappa\rightarrow 0$, but their ratio remains finite and we have $p_{\rm ex}/u_{\rm ex}=1$. The contribution from the neutralizing medium  to the excess energy (in the linear approximation) is~\cite{KhrapakPRE02_2015,HamaguchiJCP1996}
\begin{displaymath}
u_{\rm m}= -\frac{3\Gamma}{2\kappa^2}-\frac{\kappa\Gamma}{2}.
\end{displaymath}
Similarly, contribution of the neutralizing medium to the excess pressure is~\cite{KhrapakPRE02_2015}
\begin{displaymath}
p_{\rm m}= - \frac{3\Gamma}{2\kappa^2}.
\end{displaymath}
Adding these contributions we get the familiar results for the OCP within the ISM model: $u_{\rm ex}\simeq -\tfrac{9}{10}\Gamma$ and $p_0\simeq -\tfrac{3}{10}\Gamma$, which implies $p_{\rm ex}/u_{\rm ex}=1/3$. This consideration demonstrates that Yukawa systems in the limit $\kappa\rightarrow 0$ are not fully equivalent to the Coulomb (OCP) systems with the neutralizing background. Similar observation has recently been reported in relation to 2D Yukawa fluids.~\cite{KryuchkovSM2016}  

\subsection{Density scaling exponent}

Let us now consider correlations between configurational  components of energy $U$ and pressure $W$ in more detail. The density scaling exponent can be defined as~\cite{HummelPRB2015}
\begin{equation}
\gamma=\frac{\left(\partial W/\partial T\right)_V}{\left(\partial U/\partial T\right)_V}.
\end{equation}   
Substituting $W$ and $U$ and making use of the identity $T/\partial T=-\Gamma/\partial \Gamma$ the density scaling exponent becomes
\begin{equation}\label{gamma}
\gamma=\frac{p_{\rm ex}-\Gamma(\partial p_{\rm ex}/\partial \Gamma)}{u_{\rm ex}-\Gamma(\partial u_{\rm ex}/ \partial \Gamma)}.
\end{equation}
When substituting expressions for $u_{\rm ex}$ and $p_{\rm ex}$ into Eq.~(\ref{gamma}), the terms linear in $\Gamma$ will cancel out and a very simple result is obtained
\begin{equation}
\gamma=\frac{1}{3}f_{\rm Z}(\alpha\kappa).
\end{equation}
This simple expression agrees with the expected behaviour. In the limit $\kappa\rightarrow 0$ we get the expected OCP limiting value $\gamma = 1/3$, corresponding to the unscreened Coulomb interaction. For the ``Veldhorst state point'' with $\kappa=4.30$ and $\Gamma=4336.3$ (using the definitions of $\kappa$ and $\Gamma$ adopted in this paper) Eq.~(\ref{gamma}) yields $\gamma= 2.07$ in good agreement with the result obtained from a direct MD simulation,~\cite{VeldhorstPoP2015} $\gamma = 2.12$.  

Let us also consider another possible derivation of the density scaling exponent $\gamma$. For an arbitrary potential $V(r)$ an effective IPL exponent (or inverse effective softness parameter) can be introduced using ratios of derivatives of the potential,~\cite{VeldhorstPoP2015,BaileyJCP2008}
\begin{equation}\label{DSE}
n_{\rm eff}^{(p)}=-\Delta \frac{V^{(p+1)}(\Delta)}{V^{(p)}(\Delta)}-p,
\end{equation}
where $V^{(p)}$ is the $p$-th derivative of the potential, and $\Delta$ characterizes mean separation between the particles. For IPL potentials, $V(r)\propto r^{-n}$, we get $n_{\rm eff}^{(p)}\equiv n$ for any $p$ and $\Delta$. Moreover, for IPL potentials the density scaling exponent is trivially related $n$: $\gamma = n/3$ (in 3D). For other potentials, the effective IPL exponent will generally depend on $p$ and also on the exact definition of $\Delta$. Previously, $\Delta=\rho^{-1/3}$ with $p=0$ and $p=1$ were used to identify universalities in melting and freezing curves of various simple systems (Yukawa, IPL, Lennard-Jones, generalized Lennard-Jones, Gaussian Core Model, etc.).~\cite{KhrapakPRL2009,KhrapakJCP2011} It was, however, argued that the choice $p=2$ is more physically justified.~\cite{VeldhorstPoP2015,BaileyJCP2008} Indeed, it is straightforward to verify that, for the Yukawa potential, Eq.~(\ref{DSE}) with $p=2$ yields $n_{\rm eff}^{(2)}=f_{\rm Z}(\alpha\kappa)$, that is $\gamma=n_{\rm eff}^{(2)}/3$, similarly to the conventional IPL result. Thus, {\it identical} results for the density scaling exponent $\gamma$ can be obtained using the two seemingly very different routes: (i) thermodynamic approach based on explicit knowledge of the system pressure and internal energy and (ii) effective IPL exponent consideration, which operates only with the third and second derivatives of the interaction potential evaluated at the mean interparticle separation. An interesting related question, whether this is a special property of the Yukawa
interaction or perhaps a more general result, requires careful consideration and will not be discussed here.               

\subsection{Gr\"uneisen parameter}

Because the density scaling exponent does not depend on the temperature, the Gr\"uneisen parameter can be easily expressed using $\gamma$ as:
\begin{equation}\label{Grun}
\gamma_{\rm G}= \frac{1}{c_{\rm V}}\left[1+\gamma(c_{\rm V}-3/2)\right],
\end{equation}  
where $c_{\rm V} = C_{\rm V}/N$ is the reduced heat capacity at constant volume. The derivation is straightforward, for details see e.g. Ref.~\onlinecite{SchroderJCP2009}.
  
\begin{figure}
\includegraphics[width=8 cm]{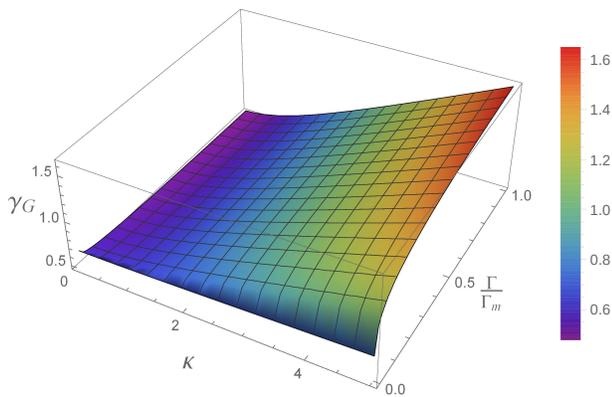}
\caption{Plot of the Gr\"uneisen gamma parameter, $\gamma_{\rm G}$, on the plane ($\kappa$, $\Gamma/\Gamma_{\rm m}$).}
\label{Fig2}
\end{figure}

\begin{figure}
\includegraphics[width=7.5cm]{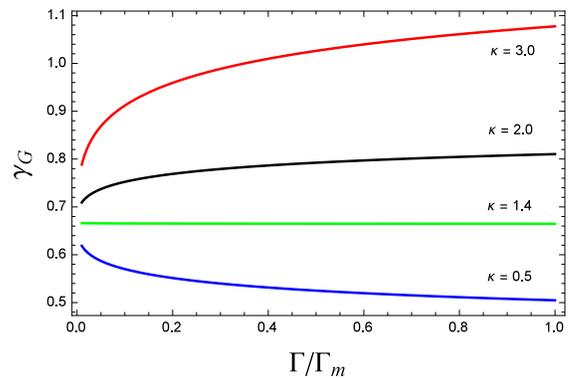}
\caption{Dependence of the Gr\"uneisen parameter $\gamma_{\rm G}$ on the reduced coupling strength $\Gamma/\Gamma_{\rm m}$ for different screening parameters, $\kappa = 0.5$, $1.4$, $2.0$, and $3.0$ (curves from bottom to top).}
\label{Fig3}
\end{figure}

The  Gr\"uneisen parameter evaluated using Eq.~(\ref{Grun}) is plotted in Figure \ref{Fig2}. Clearly, $\gamma_{\rm G}$ is {\it not} independent of temperature. Let us discuss the main trends observed.
In the limit of very weak coupling (ideal gas limit) we have $c_{\rm V}=3/2$ and hence $\gamma_{\rm G} = 2/3$, as expected for the ideal gas in 3D.~\cite{Arp1984} As the coupling becomes stronger, we can apply the RT scaling to get $c_{\rm V}\simeq 3/2+(3\delta/5)(\Gamma/\Gamma_{\rm m})^{2/5}$. Assuming that the ideal gas contribution to $c_{\rm V}$ exceeds that due to strong coupling effects (this is justified for $\Gamma\lesssim 0.5\Gamma_{\rm m}$), the following estimate is obtained
\begin{displaymath}
\gamma_{\rm G}\simeq \frac{2}{3}+\frac{6\gamma-4}{15}\delta\left(\frac{\Gamma}{\Gamma_{\rm m}}\right)^{2/5}.
\end{displaymath}  
This expression indicates that $\gamma_{\rm G}$ can either increase or decrease compared to the ideal gas value of $2/3$. The bifurcation occurs at $\gamma=2/3$, that is at $\kappa\simeq 1.4$ for Yukawa systems. 
This behaviour is further illustrated in Fig.~\ref{Fig3}, which shows the dependence of $\gamma_{\rm G}$ on the reduced coupling strength $\Gamma/\Gamma_{\rm m}$ [calculated from Eq.~(\ref{Grun})] for four different screening parameters. In particular, Fig.~\ref{Fig3} documents the existence of a range of screening parameters near the transitional value $\kappa\simeq 1.4$, where the Gr\"uneisen parameter remains close to its ideal-gas limiting value even in the strongly coupled regime. For $\kappa\gtrsim 1.4$ the Gr\"uneisen parameter increases with coupling, for  $\kappa\lesssim 1.4$
the tendency is opposite.


On approaching the fluid-solid phase transition from the fluid side, $c_{\rm V}$ reaches values slightly above $3$.~\cite{VaulinaPRE2010} In the OCP limit, the accurate analytical EoS~\cite{KhrapakCPP2016} predicts $c_{\rm V}\simeq 3.4$.~\cite{KhrapakPoP2015} The same estimate is obtained using the RT scaling (with $\delta=3.1$, as adopted here). This corresponds to the following approximation of $\gamma_{\rm G}$ for 3D Yukawa melts:
\begin{equation}\label{melts}
\gamma_{\rm G}^{\rm m}\simeq 0.56\gamma + 0.29.
\end{equation}    
The minimum value of $\gamma_{\rm G}^{\rm m}\simeq 0.48$ occurs in the OCP limit with $\kappa\rightarrow 0$ and $\gamma\rightarrow 1/3$. As $\kappa$ increases, the density scaling exponent also increases monotonously and so does the  Gr\"uneisen parameter, see Fig.~\ref{Fig3}. Finally, deep into the solid phase, the harmonic approximation is appropriate and we have $c_{\rm V}\simeq 3$ (Dulong-Petit law). In this regime $\gamma_{\rm G}^{\rm s}\simeq \gamma/2+1/3$, comparable to the result for Yukawa melt, Eq.~(\ref{melts}).

\section{Conclusion}

In this paper simple analytical expressions for the density scaling exponent and the Gr\"uneisen parameter of strongly coupled Yukawa fluids in three dimensions have been derived and analysed. It turns out that identical results for the density scaling exponent $\gamma$ can be obtained using the thermodynamic approach (based on explicit knowledge of the system pressure and internal energy) as well as from an effective IPL exponent consideration (which requires only the third and second derivatives of the interaction potential, evaluated at the mean interparticle separation). 

The Gr\"uneisen parameter evaluated here can potentially be useful in the context of shock-waves experiments in complex (dusty) plasmas. It appears in the expressions relating the pressure and density jumps across a shock wave front (known as Hugoniot equations). For a relevant example of experimental analysis and previous estimate of the Gr\"uneisen gamma the reader is referred to Ref.~\onlinecite{UsachevNJP2014}. 

The results obtained can be useful provided (i) shock-waves are excited in three dimensional particle clouds, (ii) the Yukawa potential is a reasonable representation of the actual interactions between the charged particles under these conditions, (iii) there is no or weak dependence of particle charge on particle density (in the theory described here the particle charge is constant), and (iv) the screening length is not very much smaller compared to the mean interparticle separation. These conditions can (at least partially) be met in complex plasma experiments under microgravity conditions, e.g. in the PK 4 laboratory, currently operational onboard the International Space Station.   

\acknowledgments

This work was supported by the A*MIDEX project (Nr.~ANR-11-IDEX-0001-02) funded by the French Government ``Investissements d'Avenir'' program managed by the French National Research Agency (ANR).

\bibliographystyle{aipnum4-1}
\bibliography{Gruneisen_Ref}

\end{document}